\documentclass[aps,preprint,onecolumn,nofootinbib]{revtex4}
\usepackage{tipa,mathrsfs}
\usepackage{bm}
\usepackage{mathrsfs}
\usepackage{amssymb}
\usepackage{amsmath}
\usepackage{amsfonts,dsfont}
\usepackage{color}
\usepackage{ulem}

\usepackage{graphicx}
\usepackage{amssymb}
\usepackage{amsmath}
\usepackage{amsfonts}
\usepackage{mathrsfs}
\usepackage{color}
\usepackage{ulem}
\usepackage{physics}
\usepackage[english]{babel}
\usepackage[utf8]{inputenc}
\usepackage[T1]{fontenc}
\usepackage{mathptmx}
\usepackage{etoolbox}
\usepackage[]{graphicx}
\begin{document}

\newcommand{\half}{\frac{1}{2}}
\newcommand{\pcl}[1]{#1_{\mathrm{p}}}
\title{A time (anti)symmetric approach to the double solution theory}
\author{Pierre Jamet $^{1}$ and Aur\'elien Drezet $^{2}$}
\address{$^{1}$\quad Laboratoire de Physique Subatomique et Cosmologie, UMR 5821, CNRS-Universit\'e Grenoble Alpes, 53 avenue des Martyrs, 38000 Grenoble, France\\
$^{2}$\quad Institut N\'eel, UPR 2940, CNRS-Universit\'e Grenoble Alpes, 25 avenue des Martyrs, 38000 Grenoble, France
}

\email{aurelien.drezet@neel.cnrs.fr}
\begin{abstract}
In this work we present a new theoretical approach to interpreting and reproducing quantum mechanics using trajectory-guided wavelets.  Inspired by the 1925 work of Louis de Broglie, we demonstrate that pulses composed of a difference between a retarded wave and an advanced wave (known as antisymmetric waves) are capable of following quantum trajectories predicted by de Broglie-Bohm theory (also known as Bohmian mechanics).  Our theory reproduces the main results of orthodox quantum mechanics and, unlike Bohmian theory, is local in the Bell sense.   We show that this is linked to the superdeterminism and past-future (anti)symmetry of our theory. 
 
\end{abstract}

\maketitle
 \section{Introduction}

 Following the success of his idea of wave-particle duality at the core of his quantum-wave-mechanics\cite{debroglie1923a,debroglie1923b,debroglie1923c,debroglie1924,debroglie1925th}, Louis de Broglie suggested that this notion should be understood, not as separate waves and particles connected through some constraint, but rather as a particle being itself a subcomponent of a more fundamental undulatory phenomenon. He presented this new approach, which he called the ``double solution theory'' research program, in a publication in 1927 \cite{Broglie2}. Soon however, due to the mathematical complexity of the double solution, he presented at the 1927 Solvay conference an alternative approach namely the pilot-wave theory, receiving mixed reactions overall but still claiming the interest of some of the attendees~\cite{Solvay}. History, however, showed that the Copenhagen interpretation of quantum mechanics would become the standard theory of the quantum world, and most of these ``classical'' ideas were abandoned. Even those that survived or were later revived, mostly revolved around pilot-wave theories, such as is the case with the most known hidden variable Bohmian mechanics (also named de Broglie-Bohm theory \cite{debroglie1930,Bohm1952,Bohm}, here after dBB theory), meaning that waves and particles would long remain distinct objects, even in the eyes of the most critical physicists.

 However, de Broglie left us with some pioneering works on the subject, which he first studied around the end of the 1920s~\cite{Broglie2} and then revived some decades later~\cite{Broglie3}. In particular in 1925 he briefly considered, as we will do here, the possibility of time-symmetric waves mixing retarded and advanced waves \cite{debroglie1925,debroglie1925b}, but quickly favored some non-linear dynamics as a model of particles within waves \cite{Broglie3}. In a recent work one of us developed a version of the double solution theory using the time-symmetric description and able to reproduce the main properties of quantum mechanics \cite{Drezet2023,Drezet2024}. This approach, which has some similarities with the theory of action-at-a-distance proposed by Wheeler and Feynman \cite{Wheeler,Wheelerb}, implies the existence of time-symmetric divergent fields in the vicinity of the particles. These are treated as singularities of the field, consisting of a half-sum of the retarded field emitted by the particle and the advanced field absorbed by the particle.  Note that the field singularity is then eliminated from this theory by introducing a non-linearity into the field equations that allows the field value to be fixed at a high but finite value in the central region of the particle. This type of object constitutes what is known as a soliton modeling a quantum particle.\\
 \indent  However, it is possible to develop another approach, also inspired by de Broglie's work of 1925 \cite{debroglie1925,debroglie1925b}, but which instead of using a half-sum of retarded and advanced waves $\frac{u_{ret.}+u_{adv.}}{2}$, uses a half-difference $\frac{u_{ret.}-u_{adv.}}{2}$. The advantage of using such a difference is that it allows us to remove the singular properties of the field without requiring nonlinear wave equations. This will be the subject of the present work.

 We will thus begin Section \ref{section2} with a short historical review of the main ideas behind the double solution theory, which we will follow with the general equations of our approach.
 In Section \ref{section3}, we will solve these equations in several special cases to illustrate the types of solutions we can expect from this theory. This will remain quite qualitative, focusing more on the general aspects of solutions rather than their rigorous mathematical expressions.
 Section \ref{section4} will present a further generalization of our equations recovering dBB theory. More precisely we will show  that our $u-$wavelet associated with particles can be easily guided by a dBB quantum trajectory. Therefore, since dBB theory is empirically equivalent to the usual quantum approach (as shown for example in \cite{Bohm}) the present framework allows us to recover many predictions of the standard quantum theory!  In particular, we will consider the impact that our fundamentally local theory of a $u(x)$ wave moving through spacetime has on causality in relation to Bell's theorem. We'll show that our approach can be seen as ``superdeterministic'' (for discussions of superdeterminism in relation with Bell's theorem see  \cite{Hooft1,Hooft2,Vervoort,Belltheorem,Palmer,Ciepielewski} and \cite{Drezet2023,Drezet2024}) and that it allows dBB nonlocality to emerge as a valid result at the level of Bohmian trajectories, \textit{i.e.} when we disregard the fundamental $u$ field bathing spacetime, and which is fundamentally local.
 We will close this article with a broad discussion and conclusion (see Section \ref{section5}) on some important or remarkable aspects of our work, its implications, and a proposal that it could prove very useful even outside quantum mechanics in order to develop a new generation of optical or acoustical beams and pulses guided by trajectories. 
   
\section{General framework}\label{section2}
\subsection{The main ideas of the double solution theory}\label{section21}
It is useful to recall where the idea of wave-particle duality comes from, and how it evolved into both pilot-wave theories as well as the less-known double solution program. In his 1924 Ph.D. thesis~\cite{debroglie1925th}, french physicist Louis de Broglie famously derived equations for what is refered to as ``wave-particle duality'': the idea that to each particle (electrons, photons, \dots) is associated a wave, so that Nature displays both corpuscular and undulatory phenomena in the microscopic world. Although his eponymous wavelength is now one of the starting points of most quantum mechanics courses and thus remains quite famous, the more general picture that de Broglie developed is often forgotten. In fact, even the expression for this wavelength is only part of a more general (relativistic) identity between waves and particles.

The reason we put this much emphasis on wave-particle duality, is because one can use this single fact to derive intuitively most of the basic quantum phenomena, and even some more complex ones. We can then wonder, if history had taken us on a much different course, if the theory of the quantum could have been based on these notions rather than wave-functions, matrices and probabilities. The goal is of course not to throw away the modern formalism of quantum mechanics: it works quite well as it is and another theory would not add very much. We hope, however, that by studying various approaches we can pinpoint some of the more essential aspects of quantum mechanics and gain a better understanding of this famously strange theory.

In 1924, Louis de Broglie asks how a microscopic particle like an electron ``percieves'' time,
and more importantly how it can synchronize itself with the surrounding world. This is obviously very close in spirit to Einstein's explanation of special relativity, and in fact the starting point is to assign to the particle -- for example an electron -- an ``internal clock'' whose frequency $\omega_0$ is given by the Planck-Einstein quantum formula $\hbar\omega_0 = E_0$ with $E_0$ the rest energy of the particle, so that we get
\begin{equation}
  \omega_0 = \frac{m_0 c^2}{\hbar},
\end{equation}
with $m_0$ the proper (or rest) mass of the particle. By studying how this frequency is modified in various moving frames, de Broglie shows that it must remain at all time synchronized with a superluminal phase wave of frequency $\omega = \omega_0/\sqrt{1-v^2/c^2}$, with $v$ the relative velocity between the observer and the particle. This then implies a beautiful relation between these two objects, a generalization of the Planck-Einstein relation,
\begin{equation}
  \begin{pmatrix}
    E\\
    \vb{p}
  \end{pmatrix} = \hbar\begin{pmatrix}
    \omega\\
    \vb{k}
  \end{pmatrix},
\end{equation}
with $\vb{p}$ and $\vb{k}$ the linear momentum of the particle and the wave vector of the corresponding wave respectively. This new relation $\vb{p} = \hbar\vb{k}$ is well known as de Broglie's relation, but should in fact not be separated from its 4-vector counterpart.

Importantly, this new wave is indeed a phase wave, meaning that it propagates faster than light at the velocity $v_\varphi = c^2/v$. If we want to interpret it as being a real physical object, meaning for instance that it could interact with other objects through some forces, it must obey the usual constraints of our physical world (mainly relativity), so that it is necessary to view it as only a subcomponent of a ``real'' wave in the following way:
\begin{equation}
  \text{physical wave} = \text{phase wave}\times\text{group wave}.
\end{equation}
Indeed, in any physical wave, a phase wave must always be associated with a group wave. What is remarkable is that one can show in this case that such a group wave
\begin{itemize}
  \item[i)] has the same velocity as the particle $v_g = v$.
  \item[ii)] localizes -- or more correctly, represents the spatial distribution of -- the total energy of the real wave and thus the associated particle.
\end{itemize}
It is then easy to suggest that one could describe a quantum object as being this total wave we will denote as $u$ (or being represented by it), with its wave-like properties being determined by the phase-wave, while its corpuscular aspects are themselves given by the group wave. This is the essence of the double solution theory, where a single object exists as the combination of two solutions, one for the wave and the other for the particle, but each of them inseperable from the other.
It is important to note that this is not necessarily the historical motivation behind de Broglie's theory, but this way of explaining it has the advantage of being very intuitive and tying in nicely with our subsequent derivation.

~

In previous articles~\cite{Jamet1,Jamet2,Jamet3} (written in the framework of hydrodynamic quantum analogs \cite{Couder,HQA,Fort} and `hydrodynamic quantum field' models \cite{Durey,Dagan,Darrow}), we showed that it is possible to create a basic model of a double solution using a very simple field (scalar and complex valued) $u(x)$ consisting of two counter-propagating plane waves $u_\pm$. For a 1D  problem we can write the field qualitatively as
\begin{equation}
  u(t,x) = u_+ + u_- = e^{i(\omega_+ t - k_+ x)} + e^{i(\omega_- t + k_- x)} = 2e^{i(\omega t - kx)}e^{i(\omega_g t - k_g x)}.
\end{equation}
This works on the basis that such a wave appears as a standing wave in the proper frame of the particle, which was also one of de Broglie's starting points \cite{debroglie1925,debroglie1925b}. The fact that this solution, extracted from a purely classical model, manages to account for many of the phenomena that are usually seen today as manifestations of the quantum nature of microscopic physics rather than something more general, was remarkable and very promising. There were, of course, some issues that were discussed concerning the incompleteness of the solutions ($l = 0$ orbitals in the atomic model for instance), as well as a requirement that these physical waves must display some form of superdeterminism. This last point is particularly important as it can be used to motivate the new approach we present here. Indeed, although we will not discuss it in details before Section \ref{section4}, we should already mention that Bell's theorem requires that some tradeoff be made in any theory attempting to reproduce quantum mechanics.

\subsection{An ``acausal'' approach to wave-particle duality}\label{section22}
As we have said, two simple counterpropagating waves are sufficient as a basic model of a double solution theory. Instead of writing them as propagating in opposite spatial directions, it is completely equivalent to use two waves in opposite temporal directions. The cost of doing so is obviously to remove the hypothesis of past-to-future causality in our theory, but we should remember that this is not really a drawback but more of a tradeoff, since in doing so we are also removing  any form of `magical'and conspiratorial superdeterminism. In other words, as we will see in Section \ref{section4}, the superdeterminism present in our theory is more an illusion linked to an inadequate description of the problem (\textit{i.e.} a traditional description going from the past to the future). Of course, any Cauchy problem can be translated into such a causal description, as we'll see, but in the context of our theory it's much more natural to consider the $u$ field as a whole-unit in space-time, with some signals corresponding to retarded waves and others to advanced waves.

We stress that the idea of half-advanced half-retarded waves has been used in the past, most notably by Wheeler and Feynman in their absorber theory \cite{Wheeler,Wheelerb}, but also earlier than that with the complete Li\'enard-Wiechert potentials.

The macroscopic world as we experience it is, of course, causal, and we often impose this constraint on our theories. Another way of thinking is to argue that some thermodynamical phenomenon or other will suppress retrocausal solutions.
The emergence of irreversibility from time-symmetric physics is itself a very crucial question, so that we would do it a disservice by adding a causality hypothesis by hand at any point in our work. Let us then set aside this problem, at least for this work, and focus on the very interesting dynamics that can arise from time-symmetric equations and solutions. 

We take a d'Alembert equation with a source term $\mathcal{J}(x)$ using covariant notation in flat space-time
\begin{equation}
  \partial^2 u(x) = \mathcal{J}(x)
\end{equation}
with $x$ the position 4-vector (in the following we use the convention $c=\hbar=1$ and use the Minkowski metric with signature $(+,-,-,-)$). We can write general solutions of this equation as
\begin{equation}
  u(x) = u^{(0)}(x) + \int\dd[4]{x}G^{(0)}(x;x_0)\mathcal{J}(x_0),
  \label{eq:field_sol_green}
\end{equation}
where $G^{(0)}(x;x_0)$ is a Green function solution of
\begin{equation}
  \partial^2G^{(0)}(x;x_0) = \delta^4(x-x_0).
\end{equation}
Solutions of this problem are well known and easy to obtain, and in particular we get in momentum space $k$ 
\begin{equation}
  \tilde{G}^{(0)}(k) = -\frac{1}{k^2}
\end{equation}
which gives rise to two completely equivalent solutions indexed $ret./adv.$ (or sometimes $\pm$) in real space
\begin{equation}
  G^{(0)}_{ret./adv.}(t,t_0;\vb{r},\vb{r}_0) = \frac{\delta(t - t_0 \mp \abs{\vb{r}-\vb{r}_0})}{4\pi\abs{\vb{r}-\vb{r}_0}}.
  \label{retadvvide}
\end{equation}
These two solutions characterize two different propagation directions in time. Interestingly if we look at the corresponding field solutions $u_{ret./adv.}$, we see that their half-sum is also solution, but their half-difference is solution of the homogeneous equation
\begin{equation}
  \partial^2\pqty{\frac{u_{ret.}(x) - u_{adv.}(x)}{2}} = 0.
\end{equation}
This second approach has only rarely been considered, but we feel it is worth investigating. We will take a more in depth look at an example in the next section.

\begin{figure}
  \centering
  \includegraphics[width=.5\textwidth]{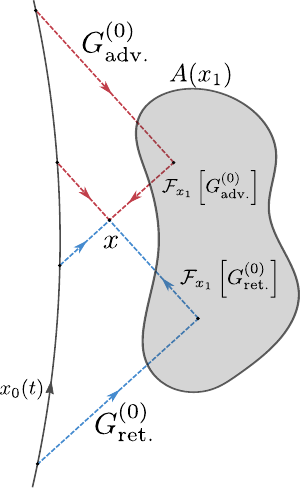}
  \caption{Space-time illustration of the total Green function seen as the sum of a direct propagation and a reflection on a potential. We want to know the field generated at position $x$ from a source following the trajectory $x_0(t)$ (the vertical axis corresponds to time $t$ and the horizontal axis to space). The retarded solutions are shown in blue, the advanced one in red. The direction of the arrows shows the direction of causal influence.  Note that while the retarded signal acts (as usual) from past to future, the advanced signal acts in the opposite direction and will be perceived by `an ideal hypothetical observer' (i.e. a kind of Maxwellian demon) perceiving time from past to future as 'superdeterministic' and 'conspiratorial'. However this is an illusion: the theory is fundamentally time-(anti)symmetric.  The gray region corresponds to the presence of a 4-potential $A_\mu$ with which the fields can interact. The field at $x$ is then given by the superposition of retarded and advanced solutions, one of each coming propagating directly from the source $x_0$, the other two having been reflected on the potential at some point $x_1$ (on which we integrate over the gray region). We clearly see the convergence of all solutions on the point $x$.}
  \label{fig:G_refl}
\end{figure}

Free solutions of the d'Alembert equations are interesting enough, but we should also consider that there could be some external potential acting upon the field. In that case, we can use a covariant derivative $D$ instead of $\partial$. For example in the case of an electromagnetic 4-potential we  have $D = \partial + i e A(x)$  (with $e$ the electric particle charge) and thus:
\begin{equation}
  D^2u(x) = (\partial + i e A(x))^2 u(x) = \mathcal{J}(x).\label{inhomoGen}
\end{equation} for the inhomogeneous wave equation 
and \begin{equation}
  D^2u(x) = (\partial + i e A(x))^2 u(x) = 0.\label{homoGen}
\end{equation} for the homogeneous wave equation
We can use the same procedure and study the Green function
\begin{equation}
  D^2G(x;x_0) = \delta^4(x-x_0).\label{Greenpropa}
\end{equation}
If we expand the covariant derivative, we get
\begin{equation}
  \partial^2G(x;x_0) = \delta^4(x-x_0) + \mathcal{F}_x G(x;x_0)
\end{equation}
with $\mathcal{F}_x$ a new operator on $G$ 
\begin{equation}
  \mathcal{F}_x = - e^2 A^2(x) - 2 i e A(x) \partial - ie (\partial A(x)).
\end{equation}
Regardless of the value of $\mathcal{F}_x$, we see that if we use the previous solution $G^{(0)}(x;x_0)$ we obtained with $A(x) = 0$, we get for the difference
\begin{equation}
  \partial^2\bqty{G(x;x_0) - G^{(0)}(x;x_0)} = \mathcal{F}_x G(x;x_0).
\end{equation}
The function $G-G^{(0)}$ can itself be solved with Green's method, so that
\begin{equation}
  G(x;x_0) = G^{(0)}(x;x_0) + \int\dd[4]{x_1}G^{(0)}(x;x_1)\mathcal{F}_{x_1}G(x_1;x_0).
\end{equation}
This equation is fundamental and tells us that the total field $G$ can be seen as the sum of a direct propagation $G^{(0)}$ and an indirect one $G^{(\mathrm{refl.})}$ corresponding to one or more (and potentially infinitely many) reflections of $G^{(0)}$ on the potential $A(x)$. This is illustrated on Figure \ref{fig:G_refl}. This is of course not something we can compute easily but, provided that $A(x)$ has a small enough amplitude, a perturbative treatement can be done. In that case, we will get $G$ as the sum of $n + 1$ terms, each representing a different number of punctual interactions with $A$, \textit{i.e.} one direct propagation plus a sum of $n$ terms associated with $i = 1,2,\dots,n$ intermediary interactions between $i+1$ propagations. This is written as
\begin{equation}
  \begin{split}
      G(x;x_0) &= G^{(0)}(x;x_0)\\
  &+ \int\dd[4]{x_1}G^{(0)}(x;x_1)\mathcal{F}_{x_1}G^{(0)}(x_1;x_0)\\
  &+ \iint\dd[4]{x_1}\dd[4]{x_2}G^{(0)}(x;x_2)\mathcal{F}_{x_2}G^{(0)}(x_2;x_1)\mathcal{F}_{x_1}G^{(0)}(x_1;x_0)\\
  &+ \dots
  \end{split}
\end{equation}

It is important to note that these solutions, both $G^{(0)}$ and $G$, are very general and can be applied to a wide range of situations, as we will see in the next section. To give back some physical meaning to these equations, let us summarize in the following way: a field $u(x)$ is generated by a (possibly point-like) source at some position $x_0$. If we choose that this source generates a total field that is expressed as the half-difference of two waves, one retarded and one advanced, the total field then behaves as if no source was present at all and we had a free field.
In general we have an inhomogeneous wave-field:
\begin{equation}
  u(x) = u^{(0)}(x) + \int\dd[4]{x}G(x;x_0)\mathcal{J}(x_0),
  \label{eq:field_sol_greenGen}
\end{equation} from which we can define a new homogeneous field as 
\begin{equation}
  u(x) = \int\dd[4]{x}\frac{G_{ret.}(x;x_0)-G_{adv.}(x;x_0)}{2}\mathcal{J}(x_0),
  \label{eq:field_sol_greenGen}
\end{equation}
 
Because of the spherical symmetry of a point-like source, we can expect the field to display the same symmetry and to decrease in amplitude as we get further from the source, similarily to the electrical field radiated from a point charge for instance. If this is indeed the case, which we will verify next, this would constitute a good candidate for a double solution. Indeed, the field can be decomposed into a product of a phase and group wave so that the increase in the amplitude of the group wave (\textit{i.e.} in the energy of the field) around $x_0$ could account for the corpuscular aspect of the field, while the phase wave could describe its undulatory nature. In this situation, the point-like particle becomes unnecessary, or rather a useful but virtual intermediary that helps us correctly derive the field.

\section{Special cases and examples}\label{section3}
There are many special cases we can study. We will choose two of them that are of particular interest in order to better understand and visualize these solutions.

\subsection{A particle at rest}\label{section31}
\begin{figure}
  \centering
  \includegraphics[width=.6\textwidth]{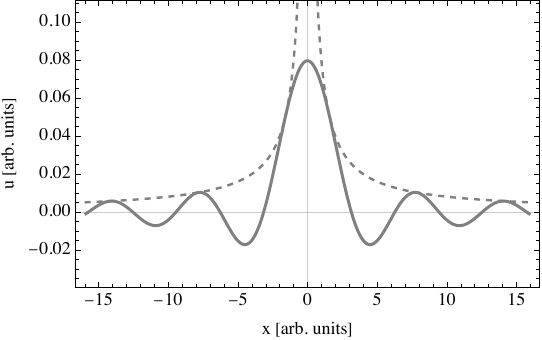}
  \caption{Qualitative representation of the basic solution for a particle at rest in arbitrary units and along one axis. A dashed outline in $1/r$ has been added to emphasize how the field energy concentrates around the source.}
  \label{fig:sinc}
\end{figure}

Let us consider a point-like particle of mass $m$ which is located at $(x_0^\mu) = (t,\vb{r}_0)$. We can choose $\vb{r}_0 = \vb{0}$ without loss of generality. This particle is also characterized by an internal periodic phenomenon $e^{-i\omega_0\tau}$ with proper frequency $\omega_0$ which we will set as the source of the field $u$. In that case, the source term can be written as
\begin{equation}
  \mathcal{J}(x) = ge^{-i\omega_0 t}\delta^3(\vb{r}),
\end{equation}
with $g$ some coupling constant. According to the equations we derived in the previous section, the total field radiated from this source particle is given in its antisymmetric form by
\begin{equation}
  u(t,\vb{r}) = \frac{g}{4\pi}\int\dd{t^\prime}\dd^3{\vb{r}^\prime}e^{-i\omega^\prime t^\prime}\frac{\delta(t - t^\prime -\abs{\vb{r}-\vb{r}^\prime})-\delta(t-t^\prime+\abs{\vb{r}-\vb{r}^\prime})}{\abs{\vb{r}-\vb{r}^\prime}}\delta^3(\vb{r}^\prime).
\end{equation}
Using the properties of the Dirac deltas, we easily see that this reduces to the simple expression
\begin{equation}
  u(t,\vb{r}) = i g \frac{\sin(\omega_0 r)}{4\pi r}e^{-i\omega_0 t}, \label{wavelet}
\end{equation}
where $r = \abs{\vb{r}}$. Some interesting comments can be made on this solution, which is drawn qualitatively on Figure~\ref{fig:sinc}.

First of all, we observe that the amplitude of the field decreases as $1/r$, which was exepected. It remains finite everywhere, even at the origin, because of the fact that we chose the half-difference of our advanced and retarded propagators. 
If we had chosen a half-sum instead, the field would have been $\cos(kr)/r$ and have diverged at 0. It is also remarkable that a solution to a linear and homogeneous wave equation can be this reminiscent of solitonic physics. Indeed, because of the equivalence of inertial frames, a source moving at a constant velocity will produce the same field, possibly contracted or stretched in some directions because of relativistic effects. In other words, a small region of space contains most of the total field energy, which can move without dissipation. This means that, according to the definition we gave in the previous section, this solution behaves in some sense both as a free wave and as a free particle. Going outside of the sole topic of quantum mechanics, this also has many implications and applications for experimental physics, optics in particular, where one could want to trap a small particle. We will discuss more of this later. An important historical point is that this particular wavelet was obtained by de Broglie as early as 1925. It describes the case of a corpuscle at rest or in uniform motion and clearly shows the possibility of seeing the quantum object described by the $u$ field as a kind of standing wave in the proper frame of reference where the particle is at rest, but that more fundamentally it is half of the difference between a retarded wave and an advanced wave.

\subsection{Qualitative discussion of an atomic model}\label{section32}
We discussed in Section \ref{section22}. how one would find the field solutions for a relativistic particle like an electron immersed in an electromagnetic 4-potential. This is a difficult thing to do properly, but we can make a few assumptions in order to get a qualitative discussion of the solutions. 

First of all, if the electron is sufficiently slow we expect relativistic effects to disappear. In the case of hydrogen, the first energy level corresponds to a velocity $v/c = \alpha \approx 0.007$, with $\alpha$ the fine structure constant. This in turn gives a Lorentz factor $\gamma = 1/\sqrt{1-v^2/c^2} \approx 1 + 2\times10^{-5}$. We recall that this factor will be used to quantify how much characteristic durations (frequencies) and distances (wavelengths) will need to be corrected because of relativistic effects. Here we get a correction of the order of $0.002~\%$, which is sufficiently small for us to neglect it most of the time, and most importantly in our current discussion.

The second diffculty concerns the presence of a Coulomb potential $V(r)$. We showed that we can use a kind of perturbative treatement to obtain the complete solution, where each additional term corresponds to an increasing number of punctual interactions. This should work in principle because of the fact that each new interaction adds a factor $\alpha$ which is of the order of $0.7~\%$ of the previous term. Once again, neglecting corrections due to the presence of this potential seems like a reasonable approximation, though it is more delicate in this case because of the additional $1/r$ factor in $V$, which suggests that we are also required to consider only points which are at large enough distances from the origin. In more general terms, we have to make the assumption of a weak field regime.

With these two elements in mind, we make the suggestion that the field generated by a point-like harmonic source orbiting in an electrostatic Coulomb potential should be, as a first approximation, that of a similar source at rest (this agrees with more general results derived in Section \ref{section41}). This means that we expect the total field to be written as
\begin{equation}
  u(t,\vb{r}) \approx i g \frac{\sin(\omega_0\vert\vb{r}-\vb{r}_0(t)\vert)}{4\pi\vert\vb{r}-\vb{r}_0(t)\vert}e^{-i\omega_0 t} \label{semic}
\end{equation}
with $\vb{r}_0(t)$ describing a circular orbit.

A more complete and rigourous derivation of these solutions in spherical coordinates can be found in \cite{Jamet4}, where a decomposition on a basis of Bessel functions $j_l$ and spherical harmonics $Y_{lm}$ was used with the following result for the field $u$:
\begin{equation}
  u(t,r,\theta,\varphi) = ig\sqrt{1 - v_n^2}e^{i L_n t}\sum_{l,m}Y_{lm}(\theta,\varphi)Y_{lm}^*(\theta_n,\varphi_n(t))\chi_n j_l(\chi_n r)j_l(\chi_n r_n).
  \label{eq:atomic_acausal_field}
\end{equation}
The various quantities are indexed $n$ when corresponding to the circualr motion of the particle on a specific quantized orbit (the model used in \cite{Jamet4} used a semiclassical approach for the trajectory corresponding  to the Bohr-Sommerfeld quantization procedure applied to the Hydrogen atom). For instance $r_n, \theta_n$ and $\varphi_n(t)$ are the spatial spherical coordinates describing the uniform circular motion of the source, $v_n$ is the instantaneous velocity, and $L_n = \gamma m v_n - E_n$ the conserved relativistic Lagrangian of the system. $L_n$ is also generally equal to the frequency of the field, but we introduce another quantity $\chi_n = \abs{L_n - m v_n / r_n}$ which acts as an effective frequency (or rather wave vector for the spatial component of the field) corrected for the presence of the central Coulomb potential.

\begin{figure}
  \centering
  \includegraphics[width=.7\textwidth]{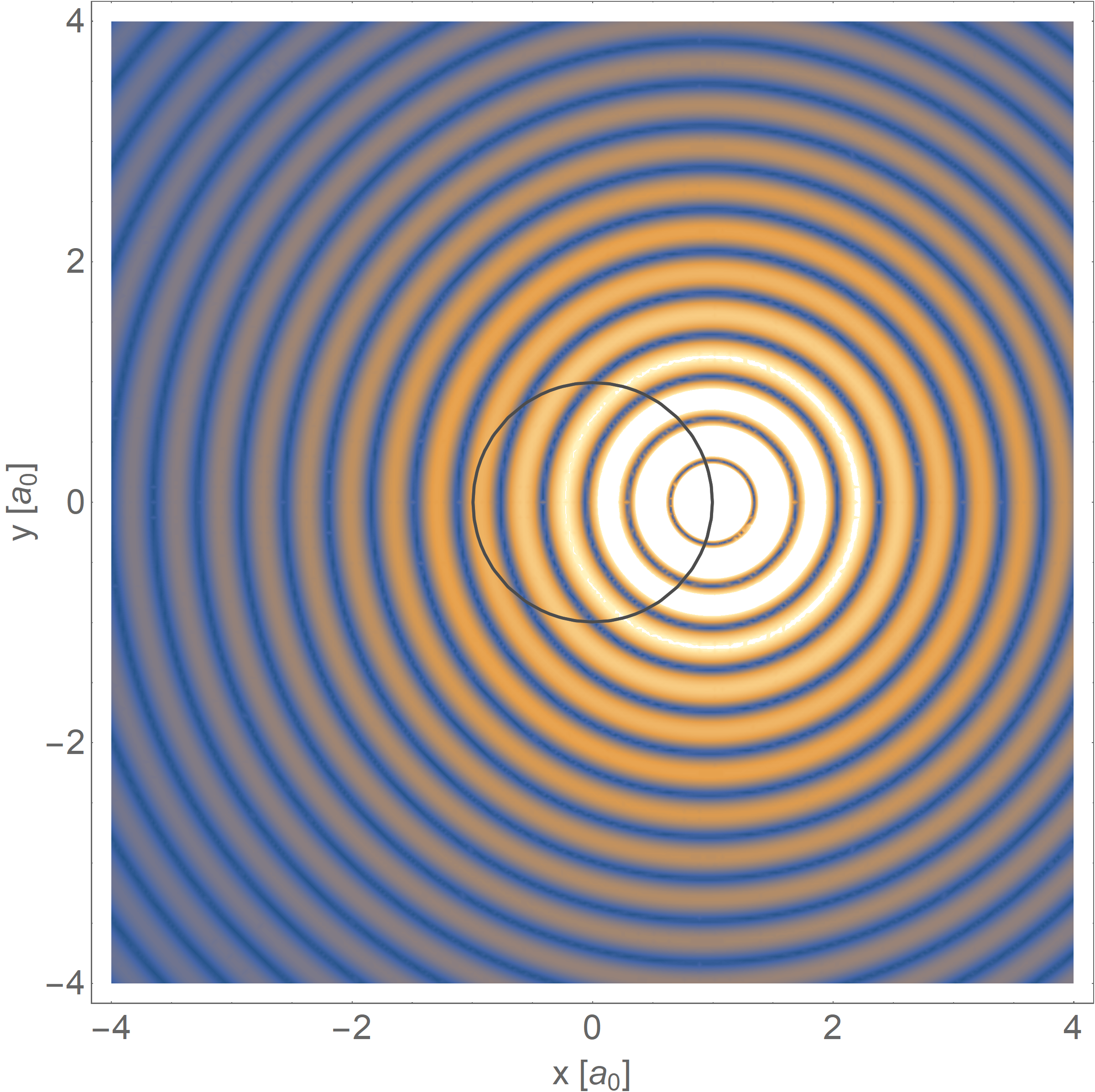}
  \caption{Numerical solution of the antisymmetric $u$-field in a central Coulomb potential in arbitrary units. The black circle corresponds to the $n = 1$ uniform circular motion of the particle in the potential well (\textit{i.e.} the electron around the nucleus). The whole $sin(r)/r$ motif moves uniformly along the orbit without changing its shape.}
  \label{fig:atom}
\end{figure}

Showing this requires lengthy calculations and the final mathematical expression is not particularily easy to interpret, which is why we do not get into much more of the details here, but interested readers are welcome to look at the full derivation (see Chap.~6, Sec.~6.2.1 and 6.3.1 of \cite{Jamet4}). This general field was then evaluated numerically as seen on Fig.~\ref{fig:atom}, and showed remarkably that the qualitative treatment in the present article is, in fact, a very good approximation, though there remain the assumptions that the electrostatic potential should not be too strong and the velocities non-relativistic.

Note that the semiclassical model is only considered here as an illustration. In a purely quantum framework, it makes more sense to use dBB dynamics, which is consistent with the results of standard quantum mechanics as far as statistical predictions are concerned. This is discussed in more detail in Section\ref{section42}.   

It's worth noting that, from a historical point of view, this type of object with a local structure in $\sin{(x)}/x$ for a field $u$ had previously been studied by Mackinnon, \cite{Mackinnon1,Mackinnon2} and by Barut \cite{Barut} and even more recently by Borghesi \cite{Borghesi} in cases of simple motions. However, the novelty of our approach lies in the possibility of generalizing in principle to any type of wavelet motion (\textit{i.e.} to cases where the equality $u=\sin{(x)}/x$ is only approximate). The key point is the use of Green's function and the half-difference between retarded and advanced solutions.

\section{Generalization}\label{section4}
\subsection{General $u$-field}\label{section41}
The preceding analysis, in particular that given in Section \ref{section2}, is in fact of a very general scope that goes far beyond the examples schematized in Section \ref{section3}. Starting from Eq.  \ref{eq:field_sol_greenGen} we see that if we choose a source term $\mathcal{J}(x)$ centered on a mean trajectory $z(\tau)$ in space-time (\textit{i.e.} in order to model a moving source) then by subtracting an advanced field from a retarded field we can define an arbitrary solution of the general homogeneous equation Eq. \ref{homoGen} imposed to follow the overall motion of the source term $\mathcal{J}(x)$. 
To be more specific, we will consider a particular source term associated with the motion of a fictitious material point moving along the trajectory $z(\lambda)$ parameterized by the affine variable $\lambda$ during motion.  We have for the source term
\begin{eqnarray}
\mathcal{J}(x)=\int_{(C)}g(z(\lambda))e^{iS(z(\lambda))}\delta^{4}(x-z(\lambda))\sqrt{(\dot{z}(\lambda)\dot{z}(\lambda))}d\lambda\nonumber\\=\int_{(C)}g(z(\tau))e^{iS(z(\tau))}\delta^{4}(x-z(\tau))d\tau\nonumber\\=g(t,\mathbf{z}(t))\delta^{3}(\mathbf{x}-\mathbf{z}(t))e^{iS(t,\mathbf{z}(t))}\sqrt{1-\mathbf{v}^2(t)}.\label{sourcepoint}
\end{eqnarray}
The first line corresponds to the most general parametrization and involves the speed $\dot{z}(\lambda):=\frac{d}{d\lambda}z(\lambda)$, the second line uses parametrization by the proper time along the path $(C)$, and the last line corresponds to the choice of parameter $\lambda=t$ \textit{i.e.} the laboratory time. We have also introduced a complex source term evolving along the trajectory $(C)$ with amplitude $g(z)$ and phase $S(z)$. Note a priori that nothing limits the speed of this fictitious point particle. The trajectory may well contain time-like segments such as $dzdz>0$ (corresponding to subluminal motion) or space-like $dzdz<0$, \textit{i.e.} tachyonic and corresponding to superluminal motion. In the following we will limit our analysis  to the more natural subluminal regime and go back to the superluminal case briefly later. 

In this formalism, the homogeneous antisymmetric field associated with the point source is written: 
\begin{eqnarray}
 u(x)=\int_{(C)}\frac{G_{ret.}(x;z(\lambda))-G_{adv.}(x;z(\lambda))}{2}g(z(\lambda))e^{iS(z(\lambda))}\sqrt{(\dot{z}(\lambda)\dot{z}(\lambda))}d\lambda\nonumber\\=\int_{(C)}\frac{G_{ret.}(x;z(\tau))-G_{adv.}(x;z(\tau))}{2}g(z(\tau))e^{iS(z(\tau))}d\tau
 \label{debroglieFFsolu}
\end{eqnarray} The crux of the problem is that here the particle with trajectory $z(\tau)$ is purely virtual in the sense that it is only a mathematical tool for finding a solution to the homogeneous equation guided by this central trajectory $z(\tau)$.  
An interesting particular solution corresponds to the material point moving in vacuum in the absence of an external field $A(x)$. We then have the Green's functions Eq. \ref{retadvvide} which after insertion in Eq. \ref{debroglieFFsolu} gives: 
\vspace{-6pt}
\begin{eqnarray}
 u(x)=\frac{1}{2}\left(\left[\frac{g(\tau)e^{iS(z(\tau))}}{4\pi\rho(\tau)}\right]_{\tau_{ret.}}-\left[\frac{g(\tau)e^{iS(z(\tau))}}{4\pi\rho(\tau)}\right]_{\tau_{adv.}}\right)\label{Lienard}
 \end{eqnarray} with $\rho(\tau)=|(x-z(\tau))\cdot \dot{z}(\tau)|$  and where the retarded proper time $\tau_{ret.}$ (respectively, advanced proper time $\tau_{adv.}$) corresponds to the point $z(\tau_{ret.})$  (respectively,  $z(\tau_{adv.})$) belonging to the trajectory $(C)$, in which $u$-radiation propagating along the forward light cone (respectively, backward light cone) is reaching the point $x$. This $u$-field is clearly reminiscent of the retarded and advanced Lienard--Wiechert potentials in classical electrodynamics and has several remarkable properties. Most importantly, near the point-particle trajectory   $x\sim z(\tau)$, \textit{i.e.},  for points located at a distance $r=\sqrt{-\xi^2}$ from the singularity in the space-like  (rest frame) hyperplane $\Sigma(\tau)$ defined by $\xi\cdot\dot{z}(\tau)=0$ (with $\xi:=x-z(\tau)$), we have approximately \cite{Drezet2023}:
\vspace{-6pt} \begin{eqnarray}
u(x)=\frac{ig(\tau)e^{iS(z(\tau))}}{4\pi}[-(\dot{S}-i\frac{\dot{g}}{g})(1+\xi\cdot\ddot{z})+\frac{i}{3}\xi \cdot z^{(3)}+O(r^2)].\label{asymp}
\end{eqnarray} where $\dot{g}:=\frac{dg(\tau)}{d\tau}$, $\ddot{z}:=\frac{d^2z(\tau)}{d^2\tau}$ and $z^{(3)}:=\frac{d^3z(\tau)}{d^3\tau}$. In particular, for a uniform motion, if $g(\tau)=const.$ and $S(\tau)=-\omega_0 \tau$ we recover the near-field of the stationary wavelet Eq.~\ref{wavelet} in the rest frame (\textit{i.e.}, we have $\frac{\sin{(\omega_0r)}}{ r}\simeq \omega_0=-\dot{S}$). This allows us to justify the local approxiamtion of Eq.~\ref{semic}. More generally, from Eq.~\ref{asymp} we deduce the field value on the trajectory:
\begin{eqnarray}
u(z(\tau))=\frac{-ig(\tau)e^{iS(z(\tau))}}{4\pi}(\dot{S}-i\frac{\dot{g}}{g})\label{asympb}
\end{eqnarray}
A second interesting property of the wavelet concerns the local phase gradient. In the particular case where $g(z(\tau))=g_0=const.$ and where the acceleration of the particle is not too large (the general case is obtained in Appendix \ref{appendix}) we have approximately 
\begin{eqnarray}
\partial^\mu\varphi(x=z(\tau))\simeq \dot{S}\dot{z}^\mu
\end{eqnarray} 
In other words the local four vector gradient of the phase is parallel to the four vector velocity of the particle. This is actually reminiscent of the  so-called de Broglie guidance formula introduced by de Broglie in his double solution theory\cite{Broglie2,Broglie3}.  We can rewrite this formula as
\begin{eqnarray}
\dot{z}^\mu\simeq\frac{\pm\partial^\mu\varphi(x=z(\tau)) }{\sqrt{[\partial^\mu\varphi(x=z(\tau)\partial_\mu\varphi(x=z(\tau)]}}
\end{eqnarray}  where the sign $\pm$ depends on the sign of $\dot{S}$. If we impose $\dot{S}<0$ (in order to recover the limit of the uniform motion where $\dot{S}=-\omega_0<0$) we have 
\begin{eqnarray}
\dot{z}^\mu\simeq\frac{-\partial^\mu\varphi(x=z(\tau)) }{\sqrt{[\partial^\mu\varphi(x=z(\tau)\partial_\mu\varphi(x=z(\tau)]}}
\end{eqnarray} or equivalently  
\begin{eqnarray}
\frac{d}{dt}\textbf{z}(t)=-\frac{\boldsymbol{\nabla}\varphi(x=z)}{\partial_t \varphi(x=z)}
\end{eqnarray}
which is the formula used by de Broglie in his  pilot wave theory  (\textit{i.e.}, Bohmian mechanics) \cite{Solvay,Bohm}.
The present analysis can be generalized to describe the motion of the wavelet in presence of an external electromagnetic field $A(x)$. Using results developed in \cite{Drezet2023,Drezet2024} for a time-symmetric soliton we can generalize  the guidance formula to   
\begin{eqnarray}
\dot{z}^\mu\simeq\frac{-(\partial^\mu\varphi(x=z(\tau))+eA^\mu(z(\tau))) }{\sqrt{[(\partial^\mu\varphi(x=z(\tau))+eA^\mu(z(\tau)))(\partial_\mu\varphi(x=z(\tau))+eA_\mu(z(\tau)))]}}\label{doublon}
\end{eqnarray} which leads to 
\begin{eqnarray}
\mathbf{v}(t)=\frac{d}{dt}\textbf{z}(t)=-\frac{\boldsymbol{\nabla}\varphi(x=z)-e\mathbf{A}(x=z)}{\partial_t \varphi(x=z)+eV(x=z)}
\end{eqnarray} again in agreement with de Broglie's double solution.\\
\subsection{Connection with Bohmian mechanics: a wavelet guided by a quantum path}\label{section42}
\indent To bring the analysis of our wavelet's motion closer to Bohmian theory, we're now going to impose that the trajectory $z(\tau)$ is a trajectory given by Bohmian mechanics for the Klein-Gordon equation. More precisely, we recall that according to quantum theory, for a scalar particle (\textit{i.e.} without spin) the relativistic wave function $\Psi(x)$ (not to be confused with our field $u(x)$) obeys the Klein-Gordon equation
\begin{equation}
  D^2\Psi(x) = (\partial + i e A(x))^2 \Psi(x) = -m^2\Psi(x)\label{KG}
\end{equation}
where $m:=\omega_0$ is the mass of the particle.  
Within the framework of the Bohmian theory, we postulate that a particle with trajectory $z(\tau)$ is guided by the wave function $\Psi$ and we demonstrate that we must have the so-called de Broglie-Bohm guiding law:
\begin{eqnarray}
\dot{z}^\mu=\frac{-(\partial^\mu S(x=z(\tau))+eA^\mu(z(\tau))) }{\sqrt{[(\partial^\mu S(x=z(\tau))+eA^\mu(z(\tau)))(\partial_\mu S(x=z(\tau))+eA_\mu(z(\tau)))]}}\label{guidanceDBB}
\end{eqnarray} 
where we use the polar expression $\Psi(x)=R(x)e^{iS(x)}$. In particular, the $S(x)$ phase is the generalization of the relativistic Hamilton-Jacobi action and obeys the equation: 
\begin{eqnarray}
(\partial^\mu S(x)+eA^\mu(x))(\partial_\mu S(x)+eA_\mu(x))=m^2+Q(x)\label{quantumpot}
\end{eqnarray} 
in which the quantum potential $Q(x)=\frac{\partial^\mu \partial_\mu R(x)}{R(x)}$ introduced by de Broglie and Bohm is involved.
We also deduce the conservation law:
\begin{eqnarray}
\partial[R^2(x)(\partial S(x)+eA(x))]=0,\label{conservation}
\end{eqnarray} linked to the local probability current conservation (see below).
In this theory the particle has a varying mass $\mathcal{M}(z(\tau))=\sqrt{[m^2+Q(z(\tau))]}$ along its trajectory (in the following we limit our analysis to the cases where $m^2+Q(x)>0$ corresponding to subluminal particle motions). The Hamilton-Jacobi action defines an integral along the trajectory:
\begin{eqnarray}
S(z(\tau))=-\int^{z(\tau)}\mathcal{M}(z(\tau'))d\tau'-e\int^{z(\tau)}A^\mu(z(\tau'))\dot{z}_\mu(\tau')d\tau'
\end{eqnarray} and we have $\mathcal{M}(z(\tau))\dot{z}^\mu(\tau)=-(\partial^\mu S(x=z(\tau))+eA^\mu(z(\tau)))$.\\
 We stress that from  Eqs. \ref{quantumpot},\ref{guidanceDBB} and \ref{conservation}, we obtain the second-order relativistic `Newton' law already found by de Broglie in 1927 \cite{debroglie1927CR}
\vspace{-6pt}
\begin{eqnarray}
\frac{d}{d\tau}[\mathcal{M}(z(\tau))\dot{z}^\mu(\tau)]=\partial^\mu [\mathcal{M}(z(\tau))]
+eF^{\mu\nu}(z(\tau))\dot{z}_{\nu}(\tau)
 \label{Newton}
\end{eqnarray}  with $F^{\mu\nu}(x)=\partial^\mu A^\nu(x)-\partial^\nu A^\mu(x)$ the Maxwell tensor field at point $x:=z$. The varying de Broglie mass $\mathcal{M}(z(\tau))$ (\textit{i.e.}, varying quantum potential $Q(z(\tau))$) is central in order to recover the non-classical features of quantum mechanics specific of the dBB pilot-wave theory.\\
\indent Returning to Eq.~\ref{conservation} we remind that the Klein-Gordon equation is associated with the local conserved current
\begin{eqnarray}
J_\mu(x):=\frac{i}{2m}(\Psi^\ast (x) D_\mu\Psi(x)-\Psi(x)D_\mu^\ast\Psi^\ast (x) )=\frac{-R^2(x)}{m}(\partial_\mu S(x)+eA_\mu(x))
\end{eqnarray}
which reads in the dBB framework $J_\mu(z(\tau))=\frac{\mathcal{M}(z(\tau))}{m}R^2(z(\tau))\dot{z}_\mu(\tau)$. In particular in the non-relativistic regime where $\mathcal{M}\sim m$ we have 
\begin{eqnarray}
J_\mu(z)\simeq R^2(z)[1,\mathbf{v}(t)]
\end{eqnarray} in agreement with Schr\"odinger equation. Indeed, we remind that in the non-relativistic limit the Klein-Gordon equation reduces to the Schr\"odinger equation. In this regime, the wave function $\Psi(t,\mathbf{x})$ obeys 
\begin{eqnarray}
i\partial_t\Psi(t,\mathbf{x})=\frac{-1}{2m}(\boldsymbol{\nabla}-ie\mathbf{A}(t,\mathbf{x}))^2 \Psi(t,\mathbf{x})+ (m+eV(t,\mathbf{x}))\Psi(t,\mathbf{x})
\end{eqnarray}
and the guiding velocity dBB becomes 
\begin{eqnarray}
\mathbf{v}(t)=\frac{d}{dt}\textbf{z}(t)=\frac{\boldsymbol{\nabla}S(x=z)-e\mathbf{A}(x=z)}{m}
\end{eqnarray} In this regime $R^2(x)=|\Psi|^2(x)$ can be interpreted as a density of probability in agreement with Born's rule. In other words if Bohmian particles are $|\Psi|^2$-distributed according to Born's law   we will be able to recover all the statistical consequences and predictions of standard quantum mechanics wih the dBB theory (at least in the non-relativistic domain). 
The $u(t,\mathbf{x})$ field associated with the wavelet defined in space-time is as follows: 
\begin{eqnarray}
 u(t,\mathbf{x})\simeq g_0\int_{(C)}\frac{G_{ret.}(t,\mathbf{x};t',\mathbf{z}(t'))-G_{adv.}(t,\mathbf{x};t',\mathbf{z}(t'))}{2}e^{iS(t',\mathbf{z}(t'))}dt'
 \label{debroglieFFsoluBohmnr}
 \end{eqnarray}
 This theory makes it possible to define wavelets following dBB trajectories, for example in the case of motion inside an atom, during a scattering process by a potential, or during a diffraction experience by a double slit giving rise to interference. 
 \begin{figure}
  \centering
  \includegraphics[width=.7\textwidth]{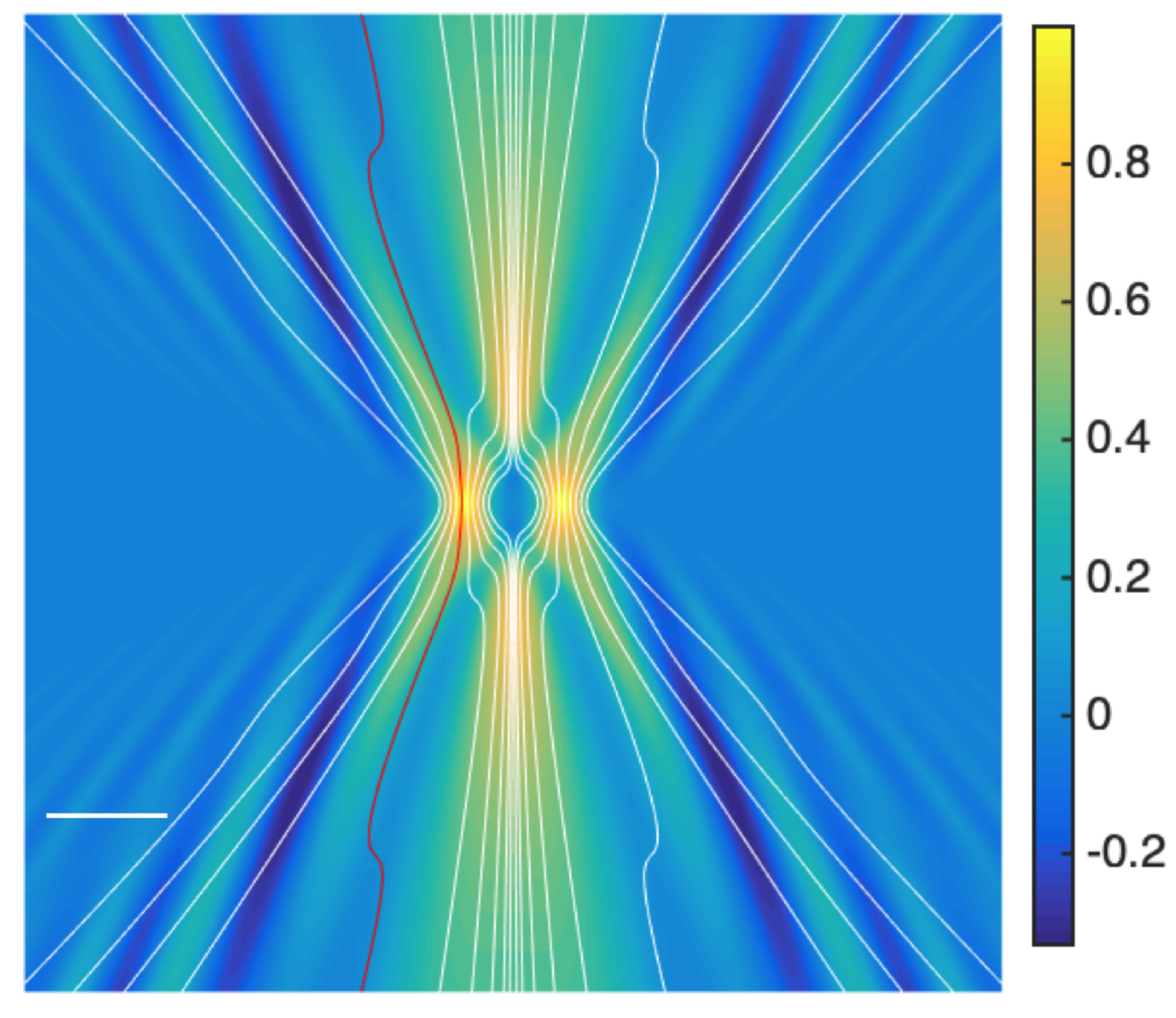}
  \caption{Non-relativistic Bohmian trajectories for a 1D Young's double slit experiment. The two Gaussian wave packets (described in the main text) generate interference patterns in the $x-t$ plane (scale bar 10 arbitrary units, here $\hbar=c=1$, mass is $m=1$ and the separation between the Gaussians is 8 units; vertical axis: time $t$ and horizontal axis space $x$).  Bohmian trajectories oscillate to reproduce interference. The red trajectory is the one used in Figure \ref{Figuredoubleslitdoublesolution}.  The color map shows the real part of the wave function (arbitrary unit).  This example has already been analyzed in the Bohmian framework in ref. \cite{Vona}.}
  \label{Figuredoubleslit}
\end{figure}
 \begin{figure}
  \centering
  \includegraphics[width=.7\textwidth]{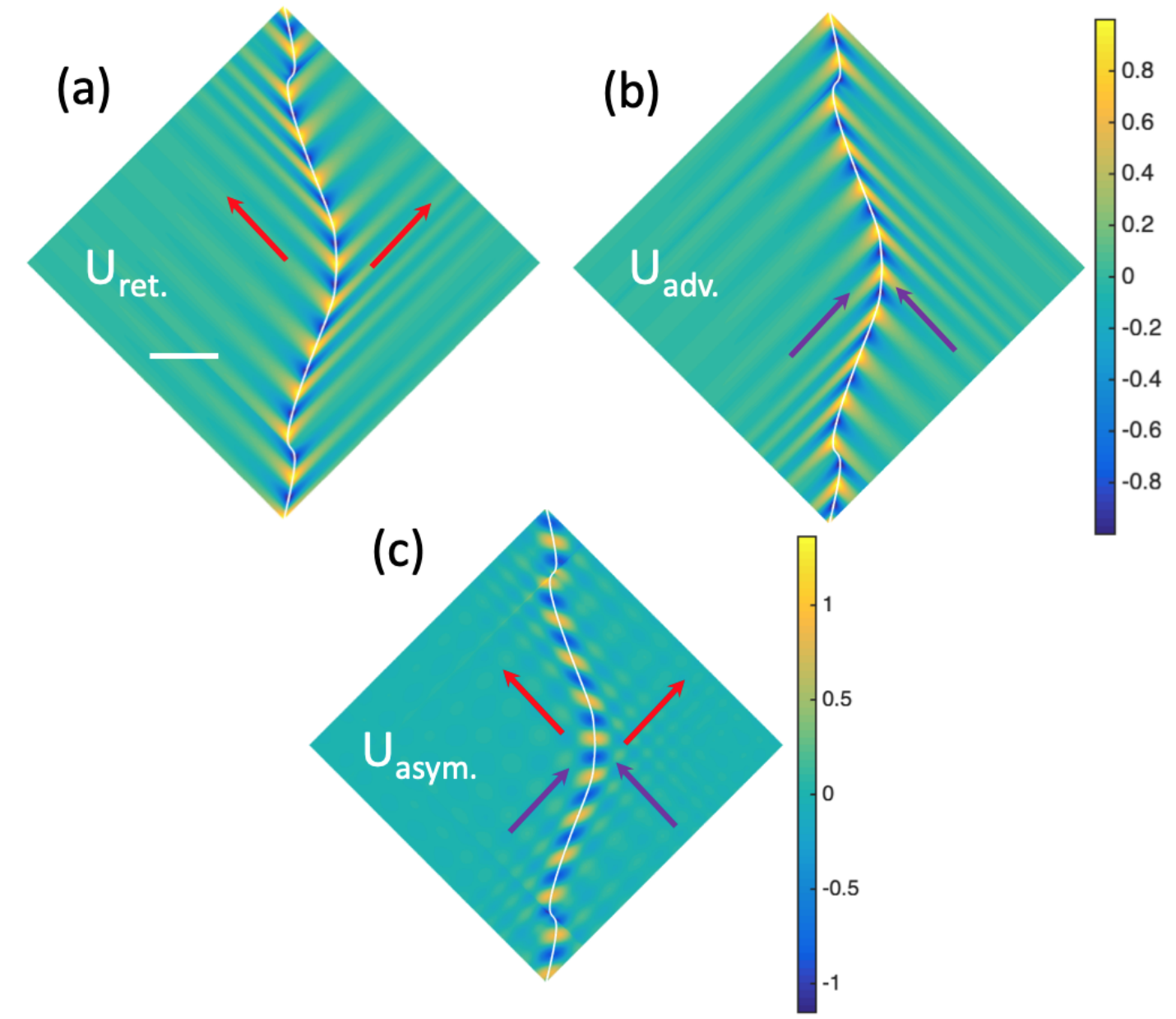}
  \caption{Field $u$ associated with a Bohmian trajectory of Young's slit experience.  The trajectory shown in red in Figure \ref{Figuredoubleslit} is used here to calculate numerically the $u(t,x,y,z)$ field in the $x-t$ plane ($x$ horizontal axis, $t$ vertical axis and the scale bar is 10 arbitrary units).  (a) shows the purely retarded field $u_{ret.}$ emitted into the future (as indicated by the red arrows) essentially along the light cones. (b) shows the advanced field $u_{adv.}$ emitted towards the past but represented here causally by blue arrows oriented in the future direction that are associated with the `conspiratorial' field originating in the past and confluent with the moving Bohmian point source. (c) shows the total field $u_{asym.}=\frac{u_{ret.}-u_{adv.}}{2}$ which recovers the Bohmian theory within the double solution framework. The diamond-shaped structure for the three images stems from computational limitations in a restricted domain of the trajectory.    The field maps show the real part of the $u$ fields considered in (a-c) (arbitrary color scale, but similar for (a) and (b)). The $1/r$ singularity for the retarded and advanced fields in (a) and (b) have been renormalized for visualization.}
  \label{Figuredoubleslitdoublesolution}
\end{figure}
It is important to note that Barut's previous approaches to diffractive phenomena could not account for the localization of the corspuscule along the trajectory (see \cite{Barut2} and a similar problem in the context of  hydrodynamic analogies \cite{Bohr}) due to the dispersion of the considered wavelets. It's worth noting that back in the 1950s, Fer \cite{Fer1,Fer2}, in collaboration with de Broglie, proposed models of particles guided by retarded waves emitted by the same particle. However, as explained in \cite{Drezet2023,Drezet2024}, these waves cannot account for the guiding formula Eq.~\ref{doublon} and are incapable of explaining interference phenomena precisely.  However, in our approach involving half-retarded and advanced fields, all these problems disappear completely. On the other hand, as we shall see, our model involving retarded and advanced waves can account for the violation of Bell's inequalities without departing from Einstein's locality. This is in fact impossible in a model involving only retarded waves.\\
 \indent As an illustration of the new theory, we consider the case of a particle described in the non-relativistic limit by a quantum superposition $\Psi(t=0,x,y,z)$ of two 1D Gaussian wave packets (along the $x$ axis) separated by a distance $d=8$ with unit variance, zero mean velocity, and initial mean position equal to -4 and +4 respectively (see \cite{Vona}). The $\Psi(t,x,y,z)$ field is distorted over time, with each Gaussian dilating towards the future and towards the past since $t=0$.  The superposition creates Young-type interferences, and the associated calculated Bohmian trajectories are shown in Figure \ref{Figuredoubleslit} in the $x-t$ plane.\\ 
 \indent The double solution theory allows us to calculate numerically the field $u(t,x,y,z)$ in the $x-t$ plane for a given Bohmian trajectory (the one shown in red in Figure \ref{Figuredoubleslit}). To do this, we use Eq. \ref{debroglieFFsoluBohmnr} with an action $S\simeq - m\tau$ (here $m=1$ and $\tau$ is the proper time).  The antisymmetric field, calculated on Figure \ref{Figuredoubleslitdoublesolution} in the $x-t$ plane, demonstrates the superdeterministic character associated with retarded and advanced wave superposition.  Note that the wavelet does follow  (\textit{i.e.}, is guided by) the Bohmian trajectory. In turn, this theory eliminates the paradoxes and limitations of previous models that do not consider advanced wavelets, and allows us to fully justify the interference  fringes (analysed in a Bohmian framework in Figure \ref{Figuredoubleslit}) using our time antisymmetric double solution theory.\:

\indent Let us now summarize the double solution theory developed here:\\
\begin{itemize}
\item[1)] We introduce a Bohmian trajectory $z(\tau)$ guided by a wave function $\Psi(z)$ solution of the Klein-Gordon Eq.~\ref{KG}.\\
\item[2)] We associate a field $u$ solution of Eqs.~\ref{inhomoGen}, \ref{sourcepoint} which is expressed in the form
\begin{eqnarray}
 u(x)=g_0\int_{(C)}\frac{G_{ret.}(x;z(\tau))-G_{adv.}(x;z(\tau))}{2}e^{iS(z(\tau))}d\tau
 \label{debroglieFFsoluBohm}
 \end{eqnarray}
and is guided by the Bohmian trajectory $z(\tau)$.\\
\item[3)] The guiding velocity of the Bohmian particle given by Eq.~\ref{guidanceDBB} is equal to the velocity of the center of the wavelet $u(x)$ moving through spacetime (see Eq.~\ref{doublon}).  
We therefore have the local equality: 
\begin{eqnarray}
\partial^\mu\varphi(x=z(\tau))\simeq \partial^\mu S(x=z(\tau))
\end{eqnarray}
where $S(z)$ is the Hamilton-Jacobi action for the Klein-Gordon equation and $\varphi(z)$ the local phase of the wave field $u(x)$ at the point $x=z(\tau)$.
\item[4)] Finally, each wavelet guided by a trajectory $z(\lambda)$ corresponds to a specific realization of dBB mechanics for a set of initial conditions $z(\lambda=0)$ and a predefined wave function $\Psi$. The statistical predictions of quantum mechanics are recovered by varying the initial conditions $z(\lambda=0)$ assumed to be distributed in accordance with Born's $|\Psi|^2$ rule (i.e. admitting the local Bohmian conservation law).
\end{itemize}

 This model and the consequences drawn from it can be generalized to systems of entangled Bohmian particles, as shown in Section \ref{section43}.
\subsection{Causality, time-(anti)symmetry and Bell's theorem}\label{section43}
\begin{figure}[h]
\includegraphics[width=10cm]{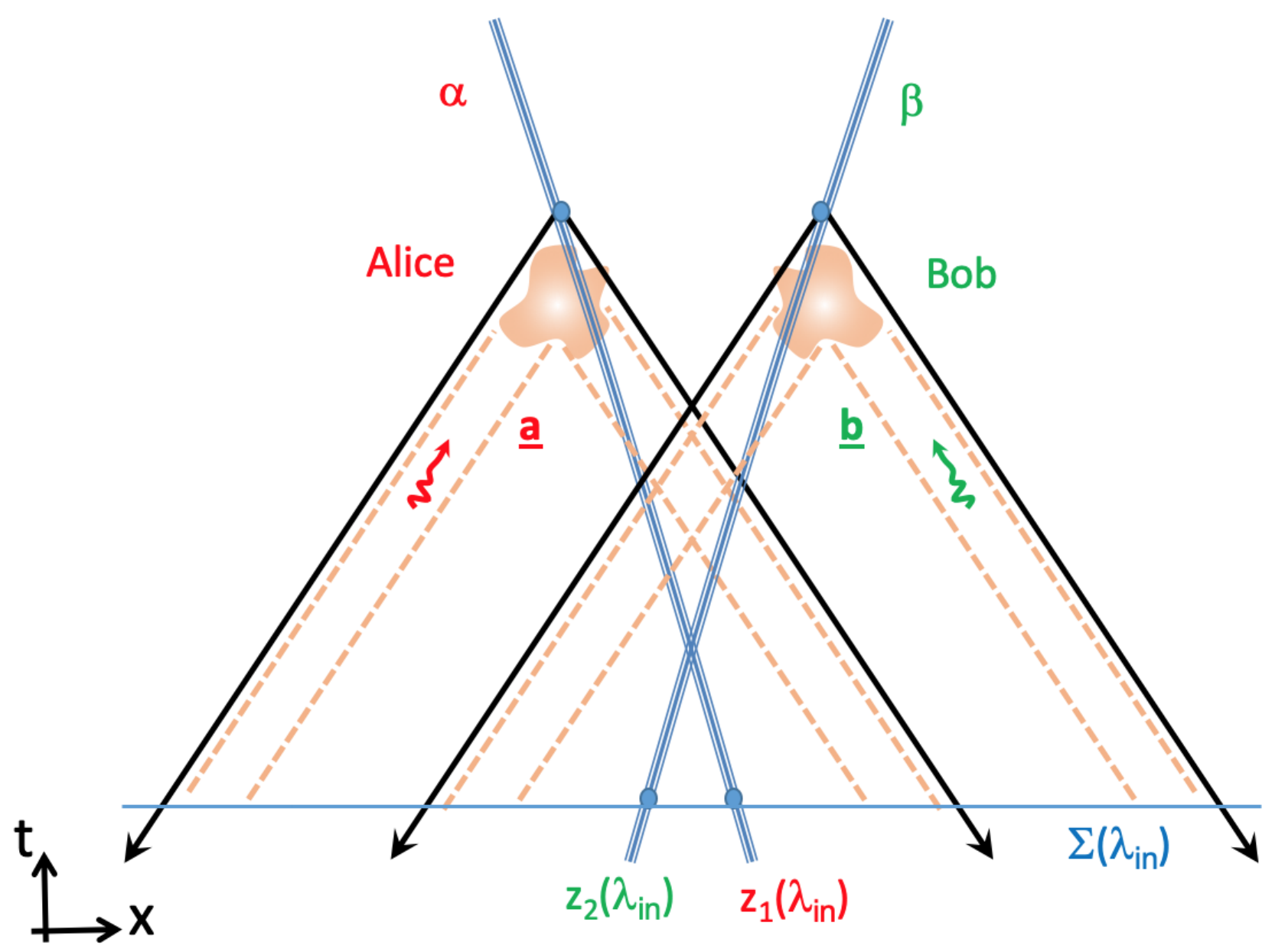} 
\caption{{Bell's} 
 scenario involving two dBB particles associated with loalized wavelets.  The information coming from the remote past and located along the space-like Cauchy hyperplane $\Sigma_{in}$ contains macroscopic classical information for changing the settings $\underline{\bold{a}}$, $\underline{\bold{b}}$ of the  external fields used by observers Alice and Bob. In the time-(anti)symmetric double solution theory, we also have advanced wave converging on the two dBB particles. These advanced waves can be seen as superdeterministic signals bringing information about the settings $\underline{\bold{a}}$, $\underline{\bold{b}}$  and located in the regions belonging to the past lights cones  with apexes in the interaction regions (in particular in the overlap of the black light cones, which is central in the derivation of Bell's theorem).  This in turn explains the  effective (but clearly not fundamental) nonlocal action-at-a-distance which is driven by our time symmetric $u$-field and which leads to Bell's correlations between observables $\alpha$ and $\beta$ associated with the two dBB particles. A similar figure was used in the context of the time-symmetric and nonlinear approach developed in \cite{Drezet2024}.  }\label{figureBell}
\end{figure}

\indent The theory developed here brings together two bodies of work that originated both in de Broglie's quest for an understanding of quantum and wave mechanics, \textit{i.e.} i) pilot wave theory aka dBB dynamics, and  ii) double solution theory.   The great specificity of our result is that it succeeds in uniting two apparently contradictory physical and ontological points of view. Indeed, the dBB theory (Bohmian mechanics) is widely known as the non-local hidden variable theory par excellence. This means that when several entangled quantum particles are considered, dBB theory predicts instantaneous action at a distance, which violates the spirit of relativity theory, but which in turn recovers quantum results and accords with Bell's theorem on nonlocality (or rather non-local-causality). On the other hand, de Broglie developed the double solution project in order to keep within a local framework in line with Einstein's relativity theory.\\  
In our approach, this duality between dBB theory and the double solution is resolved, but this can only be done at a price: that of using a causality involving retarded and advanced $u$ waves (more precisely, their half-difference $\frac{u_{ret.}-u_{adv.}}{2}$).\\  
In our opinion, the consequence is remarkable because if the theory of the double solution presented here is clearly local by applying it to an entangled particle system of the EPR (Einstein Podolsky Rosen) particle pair type, we will be able to explain the results observed in experiments of the Bell inequality violation type \cite{Bell} (like the famous Aspect \textit{et al.} experiment \cite{Aspect}) with a completely local approach!\\
To fully understand how this is possible, we need to extend our discussion of dBB theory, summarized in Subsection \ref{section42}, to a set of $N$ entangled (bosonic) particles with trajectories $\{z_j(\lambda)\}:=[z_1(\lambda),z_2(\lambda)...,z_N(\lambda)]$ and common rest mass $m=\omega_0$. We'll keep this rather schematic, as this theory is discussed in more detail in \cite{Drezet2023,Drezet2024} in the context of a different approach.  The central point is to use a multi-time formalism adapted to the Klein-Gordon equation.   In the context of the dBB pilot-wave theory, it is natural to introduce the wave function $\Psi_N(\{x_j\})=a_N(\{x_j\})e^{iS_N(\{x_j\})}$ solution of the set of $N$ coupled Klein--Gordon equations:  
\begin{eqnarray}
 D_j^2\Psi_N(\{x_j\})=-m_0^2\Psi_N(\{x_j\})
\end{eqnarray} with $D_j:=\partial_j +ieA(x_j)$ and $\partial_j$ being the 4-gradient operator for the $j${th} particle.  These equations can be deduced from a second quantization formalism but this will not be shown here. Using the polar representation, \mbox{we can write}  
\vspace{-6pt}
\begin{subequations}
\begin{eqnarray}
(\partial_j S_N(\{x_j\})+eA(x_j))^2=\omega_0^2+Q_{j}(\{x_j\}):=\mathcal{M}^2_{j}(\{x_j\}) \label{2N}\\
\partial_j[a_N^2(\{x_j\})(\partial_j S_N(\{x_j\})+eA(x_j))]=0,\label{2bN}
\end{eqnarray}
\end{subequations} with $Q_{j}(\{x_j\})=\frac{\Box_j a_N(\{x_j\})}{a_N(\{x_j\})}$ being a quantum potential.  In the context of the relativistic dBB theory, the particle velocity for the $j${th} particle is supposed to be 
\vspace{-6pt}
\begin{eqnarray}
\frac{dz_j(\lambda)}{d\lambda}=-(\partial_j S_N(\{z_j(\lambda)\})+eA(z_j(\lambda))\sqrt{\left(\frac{\dot{z}_j(\lambda)\dot{z}_j(\lambda)}{\mathcal{M}^2_{j}(\{z_j(\lambda)\})}\right)}.\label{PWIguidanceGenN}
\end{eqnarray} {Once} the initial conditions  $z_1(0),z_2(0),...,z_N(0)$ are given, this set of coupled equations can be integrated to obtain  $N$ coupled (\textit{i.e.}, entangled) trajectories for the $N$ particles. 
It is essential to note that all these trajectories are parameterized by a single $\lambda$ variable, which synchronizes the motion of the $N$ entangled corpuscles. Without this common parameter, it would be impossible to integrate the set of $N$ coupled equations of motion Eq.~\ref{PWIguidanceGenN}. This system of coupled equation generates a $N$-particle motion, which is generally strongly non-local due to the synchronization between the $\lambda$ parameter and the presence of non-local quantum potentials $Q_j(z_1(\lambda),..., z_N(\lambda))$ coupling even very distant particle.\\
\indent It is this type of nonlocality that, in dBB theory, justifies the violation of Bell's inequalities.  The relativistic dBB theory presented here in schematic form also makes it possible to recover a large number of quantum results traditionally considered inexplicable by a deterministic approach.    An essential point is that, as in the single-particle case presented in Subsection \ref{section42}, dBB theory recovers the statistical results of quantum mechanics, especially in the non-relativistic regime where we can define a probability density associated with all $N$ particles. In this non-relativistic regime, we can introduce a single common time $t:=t_1=...=t_N$ such that $\partial_t=\sum_j\partial_{t_j}$ and we deduce
\vspace{-6pt}
\begin{equation}
\partial_t a_N^2+\sum_j\boldsymbol{\nabla}_j(a_N^2\mathbf{v}_j(t))=0
\end{equation} which recovers the standard Bohmian probability law for the many-body Schr\"odinger equation with the definition $a_N:=a_N(t,\mathbf{z}_1(t),...,\mathbf{z}_N(t))$. Here,  $a_N=|\Psi_N|^2$ defines the density of probability in configuration space in agreement with Born's rule.\\
\indent Having summarized the dBB theory adapted to $N$ entangled bosons, we can now construct a completely local double solution theory. To do this, we'll start with $N$ solutions $u_j(x)$ of the linear equation Eq.~\ref{homoGen} defined in space-time and which read: 
\begin{eqnarray}
 u_j(x)=g_0\int_{(C_j)}\frac{G_{ret.}(x;z_j(\lambda))-G_{adv.}(x;z_j(\lambda))}{2}e^{iS_N(\{z_j(\lambda)\})}\sqrt{(\dot{z}_j(\lambda)\dot{z}_j(\lambda))}d\lambda
 \label{debroglieFFsoluN}
\end{eqnarray}
 The full field reads thus 
 \begin{eqnarray}
 u(x)=\sum_{j=1}^{j=N}u_j(x).
 \label{debroglieFFsoluNtot}
\end{eqnarray} 
The central point here, and we emphasize this once again, is that the $u(x)$ field is the solution of a linear, local equation defined in spacetime, whereas the Bohmian trajectories come from a highly non-local description defined in configuration space for the $N$ particles.\\  
\indent What makes all this possible is that the $u(x)$ field is the difference between a retarded and an advanced contribution. The advanced contribution in particular carries information associated with various $C_j$ trajectories in the past. This makes it possible to re-express the problem in terms of a more classical Cauchy description requiring initial conditions for the field $u(x)$ and its first derivative $\partial u(x)$ along a space-like hypersurface localized in the distant past. More precisely, applying Green's theorem to a 4-volume $\Omega$ surrounded by a closed hyper-surface (boundary $\partial \Omega$) we can express the field $u(x)$ ($x\in \Omega$) solution of Eq.~\ref{homoGen} as the integral:
\vspace{-6pt}
\begin{eqnarray}
u(x)=\oint_{\partial \Omega}\varepsilon_y d^3S_yn_y\cdot[u(y)\partial_yG(x;y)-G(x;y)\partial_y u(y)]\label{greenrete}
\end{eqnarray} where $G(x;y))$ is a Green propagator solution of Eq.~\ref{Greenpropa} ($d^3S_y$ is a three-dimensional scalar elementary volume belonging to the boundary $\partial V$ at point $y$, and  $n_y$ is the outwardly oriented normal unit four-vector at point $y$  such that $\varepsilon_y=\textrm{sign}(n_y^2)=\pm 1$). In particular, if we use the retarded  Green function $G_{ret.}(x;y)$ we can express the field at point $x:=[t,\mathbf{x}]$ as an integral over the hyperplane $t_y:=t_{in}=const.$  (with $y:=[t_y,\mathbf{y}]$) localized in the remote past of $t\gg t_{in}$ (e.g. $t_{in}\rightarrow -\infty$):
 \vspace{-6pt}
\begin{eqnarray}
u(t,\mathbf{x})=-\int d^3\mathbf{y}\cdot[u(t_{in},\mathbf{y})\partial_{t_y}G_{ret.}(t,\mathbf{x};t_{in},\mathbf{y})-G_{ret.}(t,\mathbf{x};t_{in},\mathbf{y})\partial_{t_y} u(t_{in},\mathbf{y})] \label{normalcausa}
\end{eqnarray}  where the minus sign comes from the fact that $n_y$ is time-like and past oriented (\textit{i.e.}, $\varepsilon_y=1$, $n_y:=[-1,\mathbf{0}]$).\\
\indent Moreover, from Eq.~\ref{debroglieFFsoluN} and Eq.~\ref{debroglieFFsoluNtot} the $u-$field along the hyperplane $t_{in}$ contains information about the trajectories $C_j$ coming from all time $t>t_{in}$. In other words, the initial conditions required for computing the fields $u(x)$ at any points $x$ with $t>t_{in}$ are fine-tuned, \textit{i.e.}, superdeterministic!\\
\indent The implications for our understanding of causality involving quantum systems are far-reaching.  Indeed, what we see is that starting from a traditional causal description going from the past to the future (\textit{i.e.} expressed in Eq. \ref{greenrete} with a usual retarded Green's function) we can reconstruct a set of correlated  $u-$wavelets in such a way that the associated Bohmian trajectories $z_j(\lambda)$ are non-locally entangled.  Thus, from this point of view, due to the fine-tuned initial conditions of the $u-$field, the quantum nonlocality of Bohmian trajectories is justified from a local $u$-field theory. In other words, nonlocality becomes emergent and is no longer fundamental, contrary to what Bohmians usually think!\\
\indent According to our approach, there are two levels of understanding. At the $u$-wavelet level, the description is local but superdeterministic, whereas it is nonlocal if we focus only on the motion of wavelet centers with correlated $z_j(\lambda)$ trajectories. These two levels of description are not antagonistic, however, if we recall the particular structure of $u$ fields, implying a difference between retarded (causal) and advanced (anti-causal) waves. In turn, this mixture of retarded and advanced actions explains the emergence of non-locality at the level of $z_j(\lambda)$ trajectories.\\
\indent The most important consequences concern, of course, Bell's theorem \cite{Bell,Aspect,Zeilinger,Hanson} and, more specifically, the nonlocality implied in de Broglie Bohm's theory to explain the violation of Bell's inequalities \cite{Bohm}. Let us consider a pair of entangled particles.  Although Bell's theorem is generally discussed for spin or polarization observables, we recall that for spinless particles it is possible to develop protocols for testing Bell's inequalities using dichotomous observables such as directions and momenta using interferometers (see for example \cite{Rarity} and discussion \cite{DrezetFP2019}). Without going into the details, which are unnecessary here, we'll assume a quantum state $\Psi_{1,2}$ associated with two entangled particles 1, 2 with trajectories $z_1(\lambda),z_2(\lambda)$ influencing each other non-locally at a distance (see Figure \ref{figureBell}). The two particles are separated at such distances that any subluminal communication during the experiment can be neglected. In the Bell test  Alice will observe particle 1 and Bob particle 2 and each particle is acted upon by a field (characterized by `settings' $\underline{\bold{a}}$ and $\underline{\bold{b}}$ respectively) which define the type of measurements made by Alice and Bob. 
As shown in Figure \ref{figureBell}, these settings $\underline{\bold{a}},\underline{\bold{b}}$ can in principle be determined at the last moment by Alice and Bob via two independent photons from a very distant cosmological past (e.g. emitted by two stars or quasars \cite{Zeilinger2,Zeilinger3,Kaiser,Zeilinger4}). The two Bohmian trajectories are therefore non-local functions of the settings and  of the Bohmian initial conditions, \textit{i.e.} the initial positions at a common past time $\lambda_{in}$. At a common time $\lambda_{out}$ after the measurement we have: 
\vspace{-6pt}
  \begin{eqnarray}
z_1(\lambda_{out}):=F_{1,\Psi_{12}}(\lambda_{out},z_1(\lambda_{in}),z_2(\lambda_{in}),\underline{\bold{a}},\underline{\bold{b}})\nonumber\\
z_2(\lambda_{out}):=F_{2,\Psi_{12}}(\lambda_{out},z_1(\lambda_{in}),z_2(\lambda_{in}),\underline{\bold{a}},\underline{\bold{b}}).
\label{inter12}
\end{eqnarray} 
In turn, these trajectories precisely determine the dichotomous $\alpha$ and $\beta$ observables measured by Alice and Bob in their Bell test. We therefore have the non-local functions:  
\vspace{-6pt}
\begin{eqnarray}
\alpha=A_{\Psi_{12}}(z_1(\lambda_{in}),z_2(\lambda_{in}),\underline{\bold{a}},\underline{\bold{b}}):=A_{\Psi_{12}}(\boldsymbol{\Lambda},\underline{\bold{a}},\underline{\bold{b}})\nonumber\\
\beta=B_{\Psi_{12}}(z_1(\lambda_{in}),z_2(\lambda_{in}),\underline{\bold{a}},\underline{\bold{b}}):=B_{\Psi_{12}}(\boldsymbol{\Lambda},\underline{\bold{a}},\underline{\bold{b}})\label{model1}
\end{eqnarray}  and depend nonlocally on the settings $\underline{\bold{a}},\underline{\bold{b}}$ and the local beables $\boldsymbol{\Lambda}\equiv [z_1(\lambda_{in}),z_2(\lambda_{in})]$. These are the quantities involved in the Alice and Bob joint measurements  leading to the violation of Bell inequalities. Once more, for a Bohmian, the situation is clearly demonstrating the necessary nonlocal link based on an instantaneous action-at-a-distance.\\
\indent However, from the point of view of our local wavelet theory, the total $u$ field reads 
\begin{eqnarray}
 u(x)=g_0\int_{(C_1)}\frac{G_{ret.}(x;z_1(\lambda))-G_{adv.}(x;z_1(\lambda))}{2}e^{iS_N(\{z_j(\lambda)\})}\sqrt{(\dot{z}_1(\lambda)\dot{z}_1(\lambda))}d\lambda\nonumber\\
 +g_0\int_{(C_2)}\frac{G_{ret.}(x;z_2(\lambda))-G_{adv.}(x;z_2(\lambda))}{2}e^{iS_N(\{z_j(\lambda)\})}\sqrt{(\dot{z}_2(\lambda)\dot{z}_2(\lambda))}d\lambda
 \label{debroglieFFsoluEPR}
\end{eqnarray}
 which is a function of the two Bohmian trajectories $z_1(\lambda),z_2(\lambda)$. This antisymmetrical $u$ field, made up of a retarded and an advanced part, can alternatively be seen  (see Eq.~\ref{normalcausa}) as being generated by knowledge of the $u$ field and its $\partial_t u$ derivative on a $\Sigma(\lambda_{in})$ space-like hyper-plane associated with an initial time $t_{in}$ in the distant past (see Figure \ref{figureBell}). 
\vspace{-6pt}
This initial field seems conspiratorial or super-deterministic, as it already contains, in advance, information about what will happen to the wave packets $u_1$ and $u_2$ in the future via the interaction with the settings $\underline{\bold{a}},\underline{\bold{b}}$. From this point of view, we have demonstrated that Bohmian quantum non-locality can be alternatively reproduced by a local but superdeterministic model involving retarded and advanced $u$ fields.     
\section{Summary and conclusions}\label{section5}
\indent To sum up our work, we have presented an alternative theory that is empirically equivalent to traditional quantum mechanics, based on the idea of a double solution dating back to de Broglie.   In our approach, systems of N quantum particles (entangled or not) are described by wavelets $u(x)$ moving in space-time and solutions of a linear wave equation.  The wavelets are also spatially centered on  trajectories given by Bohmian mechanical theory (or de Broglie Bohm theory). These guiding trajectories are themselves predicted by the $\Psi$ function propagating in configuration space for a set of $N$ particles.  Therefore, from the point of view of our theory, the $\Psi$ wave function is simply a tool for solving our linear wave equation for the $u$ field in space-time. \\ 
\indent Our approach reproduces the consequences of Bohmian mechanics and hence of quantum physics. Moreover, unlike Bohmian mechanics, it is entirely local in the Bell sense. In fact, Bohmian non-locality, which justifies experimentally observed violations of Bell's inequalities, emerges here as a consequence of a local theory involving retarded and advanced waves (more precisely, their half-difference).  This mixed retarded-advanced character in turn allows our theory to be interpreted as a superdeterministic approach, without having the miraculous or conspiratorial character usually associated with it. We want to insist however on the fact that this superdeterministic interpretation is only here if we insist on imposing a single time direction, in particular by reinterpreting advanced waves as being convergent retarded waves coming from a common past. Should we embrace the existence of advanced solutions with the same credence as retarded ones, then there is absolutely no need to invoke superdeterminism since the advanced waves are generated by the source itself rather than the past. In other words, an acausal (time-(anti)symmetric) phenomenon appears superdeterministic only for causal observers, but is fundamentally not.\\
\indent     It should be noted, however, that the consequences of the present work go beyond the scope of quantum mechanics and its interpretation. Indeed, from an application point of view, we have demonstrated that it is theoretically possible to generate beams guided by arbitrary $z(\tau)$ trajectories. However, in the optical or acoustic field, it has been demonstrated in the past  since the work of Durnin  et al. \cite{Durnin} that certain beams known as “focus wave modes”, “Bessel Gauss Pulses”, “X Waves” etc. \cite{Sheppard,Saari0,Hall} possess a set of properties associated with a particle or energy packet (some can even move locally faster than light \cite{Saari1,Saari}). Recent works even showed the possibility to observe some de Broglie-Mackinnon pulses \cite{Hall} which are particular cases of the waves considered here.  In the framework we're developing here, we see that we can use a clever mix of retarded and advanced waves to generate new trajectory-guided beams. Since the trajectory is not necessarily time-like, we can also imagine a new family of superluminal “X-wave” type beams: this opens up interesting prospects in communication and optical or acoustic trapping.  Finally, it also opens up fascinating prospects for developing optical or acoustic quantum analogues using `superdeterministically' pre-prepared waves (a bit like the time-reversed beam reconstruction methods developed by Fink \cite{Fink}) so as to mimic quantum predictions exactly.

\appendix
\section[\appendixname~\thesection]{Local properties of the field near the particle trajectory}
\label{appendix}
\indent For any point $x\in \Sigma(\tau)$ near $x=z(\tau)$ we have  
\begin{eqnarray}
\frac{u(x)}{u^\ast(x)}=e^{i2\varphi(x)}=e^{i2\varphi(z(\tau))}[1+2i\xi\cdot \partial\varphi(x=z(\tau))+O(r^2)]\nonumber\\=-e^{i2S}\frac{\dot{S}-i\frac{\dot{g}}{g}}{\dot{S}+i\frac{\dot{g}}{g}}[1-\frac{2i}{3}\frac{\dot{S}}{(\dot{S})^2+(\frac{\dot{g}}{g})^2}\xi\cdot z^{(3)}+O(r^2)]\nonumber\\
=-e^{i2S(z(\tau))}e^{i2\chi(z(\tau))}[1-\frac{2i}{3}\frac{\dot{S}}{(\dot{S})^2+(\frac{\dot{g}}{g})^2}\xi\cdot z^{(3)}+O(r^2)]
\end{eqnarray} with $\chi(z(\tau))=-\arctan{(\frac{\frac{\dot{g}}{g}}{\dot{S}})}+n\pi$ and $n\in \mathbb{Z}$.
From this we deduce
\begin{eqnarray}
\varphi(z(\tau))=S(z(\tau))+\pi/2+\chi(z(\tau)) +m\pi & \textrm{with }  m\in \mathbb{Z}, \label{truc1}\\
\xi\cdot \partial\varphi(x=z(\tau))=\frac{-1}{3}\frac{\dot{S}}{(\dot{S})^2+(\frac{\dot{g}}{g})^2}\xi\cdot z^{(3)}.\label{truc2}
\end{eqnarray}  From Eq.~\ref{truc1} we deduce 
\begin{eqnarray}
\frac{d}{d\tau}(S+\chi)=\dot{S}(\tau)+\dot{\chi}(\tau)=\dot{z}\cdot \partial\varphi(x=z(\tau)) \label{truc3}
\end{eqnarray}
Combining Eqs.~\ref{truc2}, \ref{truc3} we get the local phase gradient
\begin{eqnarray}
\partial^\mu\varphi(x=z(\tau))=(\dot{S}+\dot{\chi})\dot{z}^\mu-\frac{1}{3}\frac{\dot{S}}{(\dot{S})^2+(\frac{\dot{g}}{g})^2}[{z^{(3)}}^\mu-(\dot{z}\cdot z^{(3)})\dot{z}^\mu]
\end{eqnarray} In  the rest frame this reads
\begin{eqnarray}
\partial^\mu\varphi(x=z(\tau))=[\dot{S}+\dot{\chi},-\frac{\dot{S}}{3}\frac{1}{(\dot{S})^2+(\frac{\dot{g}}{g})^2}\dot{\mathbf{a}}]
\end{eqnarray} with $\dot{\mathbf{a}}:=\frac{d^3\mathbf{z}(t)}{dt^3}$ the derivative of the local acceleration in the rest frame.
In the rest frame we write $|\dot{\mathbf{a}}|\sim 1/l^2$ with $l$ a typical length characterizing the particle motion ($l\rightarrow +\infty$ if the motion is uniform) and  we also write $(\dot{S})^2+(\frac{\dot{g}}{g})^2 \sim 1/\lambda_0^2$ where $\lambda_0\sim 1/m$ is of the order of Compton's wavelength  of the particle. Now, if the movement is not too much accelerated, we have 
\begin{eqnarray}
\frac{|\dot{\mathbf{a}}|}{(\dot{S})^2+(\frac{\dot{g}}{g})^2}\sim(\frac{\lambda_0}{l})^2\ll 1 
\end{eqnarray} and therefore we can write  
\begin{eqnarray}
\partial^\mu\varphi(x=z(\tau))\simeq (\dot{S}+\dot{\chi})\dot{z}^\mu
\end{eqnarray}


\end{document}